\documentclass[a4paper,11pt]{article}
\usepackage{pos}
\usepackage{subcaption}

\def\Journal#1#2#3#4{{#1} {\bf #2}, #3 (#4)}

\title{Radio signatures of cosmic-ray showers with deep in-ice antennas}

\author*[a]{Simon Chiche}
\author[a]{Nicolas Moller}
\author[c,b]{Abby Bishop}
\author[b]{Simon de Kockere}
\author[b]{Krijn D. de Vries}
\author[b]{Uzair Latif}
\author[a]{Simona Toscano}

\affiliation[a]{Inter-University Institute For High Energies (IIHE), Université libre de Bruxelles (ULB), \\
Boulevard du Triomphe 2, 1050 Brussels, Belgium}

\affiliation[b]{Vrije Universiteit Brussel (VUB), Dienst ELEM, Pleinlaan 2, B-1050, Brussels, Belgium}

\affiliation[c]{Department of Physics, Wisconsin IceCube Astrophysics Center, University of Wisconsin-Madison,
Madison, WI 53706}
\emailAdd{simon.chiche@ulb.be}

\abstract{To detect ultra-high-energy neutrinos, experiments such as ARA and RNO-G target the radio emission these particles induce when cascading in the ice, using deep antennas in South Pole or in Greenland. One of the main backgrounds for such signals is the radio emission generated by cosmic-ray showers, either directly in the ice, or in the air and transmitted to the ice, which can both reach the deep antennas. The first detection of cosmic rays with deep antennas would thus validate this detection principle and allow us to calibrate the detectors. FAERIE, the Framework for the simulation of Air shower Emission of Radio for in-Ice Experiments, is a numerical tool that couples both CoREAS and GEANT4 Monte-Carlo codes to simulate the radio emission from cosmic-ray showers deep in the ice. Using this code, we will investigate cosmic-ray radio signatures and the possible implications on the design of a cosmic-ray veto.}

\FullConference{10th International Workshop on Acoustic and Radio EeV Neutrino Detection Activities (ARENA2024)\\
11-14 June 2024\\
The Kavli Institute for Cosmological Physics, Chicago, IL, USA\\}

\begin{document}
\maketitle

\section{Introduction}


In-ice radio detection of high-energy astroparticles is a promising technique to detect the first ultra-high-energy neutrinos ($E>10^{17}\, \rm eV$) and open a new astronomical window, thanks to the gigantic effective volume it can instrument. The method was pioneered by experiments such as RICE~\cite{Kravchenko_2003}, ANITA~\cite{Gorham_2009,Gorham_2019,Gorham_2021} and ARIANNA~\cite{Arianna,Anker_2019}, that probed its feasibility while setting limits on the neutrino flux. The interaction of a neutrino with the ice creates a cascade of secondary particles resulting in a radio emission that can reach antennas located either at the ice's surface or buried deep in the ice. ARA~\cite{Miller_2012,Allison_2016} and RNO-G~\cite{Aguilar_2021} in particular, are two in-ice experiments with close detection concepts that rely on strings of antennas buried at various depths in the ice to detect UHE neutrino. While RNO-G is still under construction, it will build on the knowledge gained from previous in-ice experiments, feature more stations than ARA, and combine both surface and deep antennas, making it one of the most sensitive next-generation UHE neutrino detectors. Yet, in-ice experiments should also detect the radio emission from cosmic-ray air showers, which can propagate into the ice and reach the deep antennas, acting as one of the main backgrounds for neutrino searches. The cosmic-ray flux is much larger than the neutrino flux, which implies that: (1) cosmic-ray detection should be more readily attainable to neutrino detection and could thus help calibrate in-ice radio detectors and validate their detection principle, (2) neutrino/cosmic-ray discrimination is needed to ensure a successful neutrino detection. In this work, we use the FAERIE Monte-Carlo code~\cite{Kockere_2024} that simulates the radio emission from cosmic-ray air showers for in-ice observers to characterize this emission and identify preliminary cosmic-ray signatures.

\section{Radio emission from cosmic-ray air showers for in-ice observers}\label{sec:CRemission}

\begin{figure}[tb]
\centering 
\includegraphics[width=0.60\linewidth]{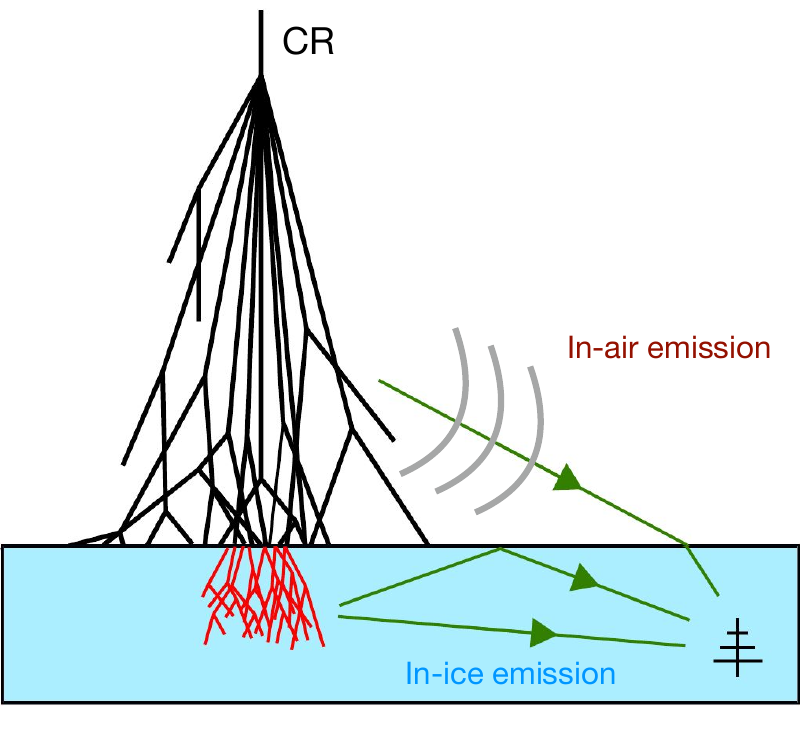}
\caption{Sketch of the in-ice radio emission from a cosmic-ray air shower. A first emission coming from the in-air cascade (represented by the black lines) is transmitted into the ice and can reach the deep antennas. A secondary emission comes from the in-ice cascade (red lines) and can also reach the deep antennas through either a direct path or a reflected emission at the air/ice boundary.}\label{fig:CRsketch}
\end{figure}

A typical sketch of the radio emission from a cosmic-ray air shower seen by an in-ice observer is displayed in Fig.~\ref{fig:CRsketch}. The emission can be divided into two parts. First, the emission from the {\it in-air cascade}: as the shower develops, radio waves emitted in the air propagate without attenuation, a part of this emission is transmitted to the ice and can reach the deep antennas. Second, the emission from the {\it in-ice cascade}: if the shower is energetic enough, some particles can penetrate the ice and induce a secondary particle cascade emitting radio waves that can also reach the deep antennas. For the emission from the in-air cascade, we expect that both geomagnetic and charge-excess mechanisms contribute to the radio signal. However, for the in-ice cascade only a charge-excess emission is expected due to the higher density of the medium in which the shower develops.  This yields different polarization signatures between the emissions from the in-air and the in-ice cascade. The geomagnetic emission is linearly polarized in the $\mathbf{v \times B}$ direction, where $\mathbf{v}$ is the  direction of propagation of the shower and $\mathbf{B}$ is the direction of the local Earth magnetic field. On the other hand the charge-excess emission is radially polarized in a plane perpendicular to the shower propagation axis~\cite{Schr_der_2017}.

The relative contribution between the in-air and the in-ice emission is highly dependent on the shower inclination (zenith angle). For vertical showers, the shower maximum $X_{\rm max}$, is reached close to the ice surface and we expect that many energetic particles can penetrate the ice and contribute to the in-ice emission. On the opposite, for inclined showers, the propagation distance between $X_{\rm max}$ and the ground is large, hence most particles lose their energy before reaching the ice which yields a higher in-air/in-ice ratio.

Additionally, the Cerenkov cone, which defines the region where the electric field amplitude is highest, is observed at different angles for in-air and in-ice emissions. The Cerenkov angle is given by $\theta_{c} = 1/\arccos{(n \beta)}$, where $n$ is the index of refraction of the medium and  where $\beta =  v/c$, with the $v$ the velocity of the shower particles. Hence, we expect a larger Cerenkov angle for the in-ice emission, with $\theta_c^{\rm air} \sim 1^{\circ}$ and $\theta_c^{\rm ice} \sim 50^{\circ}$. Eventually, both the emission from the in-air and the in-ice cascades should be bent while propagating in the ice, due to the rapidly varying density of the medium.

\section{Library of cosmic-ray showers}

To study in-ice radio signatures of cosmic-ray air showers we use the FAERIE simulation code. In this section we discuss the different steps to set up the simulations and build a library of cosmic-ray events.
\subsection{FAERIE simulation code}
FAERIE, the Framework for the simulation of Air shower Emission of Radio for in-Ice Experiments~\cite{Kockere_2024}, is a numerical tool that combines CORSIKA 7.7500~\cite{Heck:1998vt} and GEANT4~\cite{ALLISON2016186} Monte-Carlo codes to simulate both the cosmic-ray induced in-air and in-ice cascades respectively. The radio emission is then generated using the CoREAS radio extension~\cite{HuegeCOREAS} for the  in-air cascade, while a code from the T-510 experiment is used for the in-ice cascade~\cite{T510}. In both cases, the computation of the radio emission relies on the endpoint formalism~\cite{PhysRevE.84.056602}, which derives the emission from any single charge in the particle cascades  using straight-line propagation of rays. For a charged particle at a position $\Vec{x}$ moving along a track at a speed  $\Vec{\beta}^{\star} = \Vec{v}^{\star}/c$ during a time $\Delta t$, the electric field seen by an observer at a distance $R$ reads~\cite{Kockere_2024}
\begin{equation}
    E_{\pm}(\Vec{x}, t) = \pm \frac{1}{\Delta t} \frac{q}{c} \left(\frac{\widehat{r} \times [\widehat{r} \times \Vec{\beta}^{\star}]}{|1-n \Vec{\beta}^{\star} \cdot \widehat{r}|R}\right) \ ,
\end{equation}\label{eq:Efield}
where the plus/minus sign is applied to the start/end point of the track, $q$ is the particle charge, $n$ the medium refractive index and $\widehat{r}$ the unit vector that goes from the emission point to the observer.

\begin{figure}[tb]
\centering 
\includegraphics[width=0.60\linewidth]{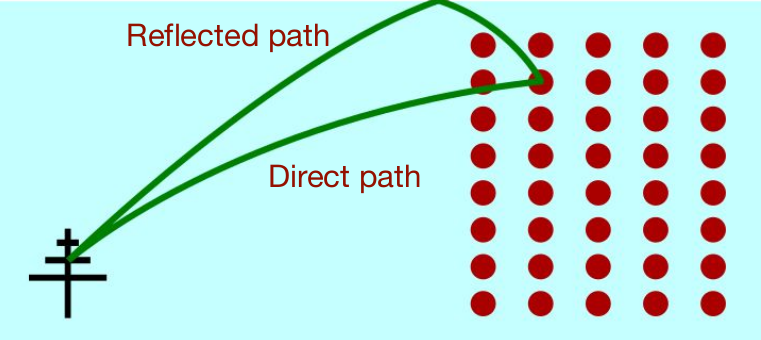}
\caption{Ray-tracing solutions (green lines), for a set of observer (antenna) and receivers (red circles). For a given observer/receiver we find two ray-tracing solutions that link the two positions: a direct path and a reflected path at the air/ice boundary.}\label{fig:RayTracing}
\end{figure}

Eventually, to account for the changing refractive index profile in both air and ice, as well as the transition of the radiation from air to ice, FAERIE relies on ray-tracing and the endpoint formalism needs to be adapted. One of the main modifications being that in the ice the variable $R$ in Eq.~\ref{eq:Efield} must be re-interpreted as the geometrical path length connecting the emitter and the receiver. This length can be derived with ray-tracing using Snell's law $n_1\sin{i_i} = n_2\sin{i_2}$, which links the incident and refracted angles $i_{1,2}$ of a given ray at the transition boundary between two medium with refractive indices $n_{1,2}$. For a given point-like emitter and receiver we find two solutions, a direct and a reflected path, as shown in Fig.~\ref{fig:RayTracing}.

\subsection{Simulation setup}

Using the FAERIE simulation code we then build a library of cosmic-ray events. To propagate the radio emission we first need to define an ice-model describing the evolution of the ice refractive index as a function of depth. The ice density can usually be well modeled using an exponential profile following $n(z) = A - B\exp{-C|z|}$, with $A$, $B$, $C$, being three fit parameters and $|z|$ the depth. For the South Pole ice we use $A =1.775$, $B =0.43$, $C =0.0132$~\cite{Kelley:2017ws}. For Greenland however, the ice profile is such that at a depth of $|z| = 14.9{\, \rm m}$ the ice reaches a critical density and becomes more compact so using a double exponential profile is more suitable. Hence we find $A = 1.775$, $B =0.5019$, $C =0.03247$ for $|z| < 14.9{\, \rm m}$ and $A = 1.775$, $B =0.448023$, $C =0.02469$ for $|z| > 14.9{\, \rm m}$~\cite{RNOice}.

For the antennas layout we consider several square grids of antennas at depths following the position of the antennas along ARA's strings and RNO-G's main triggering string (power line). For ARA we consider $12$ depths for each grid, evenly spaced between $145\, \rm m$ and $200\, \rm m$. For RNO-G we consider 5 depths $[0, 40, 60, 80, 100]\, \rm m$. To build our cosmic-ray library we generate showers for $4$ energies $E = [10^{16.5}, 10^{17}, 10^{17.5}, 10^{18}]\, \rm eV$, one azimuth angle $\varphi = 0^{\circ}$ (shower propagating towards magnetic north) and $9$ zenith angles evenly spaced in $\cos{\theta}$, $\theta = [0^{\circ}, 10^{\circ}, 20^{\circ}, 28^{\circ}, 34^{\circ}, 39^{\circ}, 43^{\circ}, 47^{\circ}, 50^{\circ}]$. 
\section{Simulation results}
Once the simulation inputs are set, we generate the library of cosmic-ray showers and discuss some of the simulation results below.
\subsection{Electric field maps}
\begin{figure}[htbp]
    \centering
    \begin{minipage}[b]{0.47\textwidth}
        \centering
        \includegraphics[width=\textwidth]{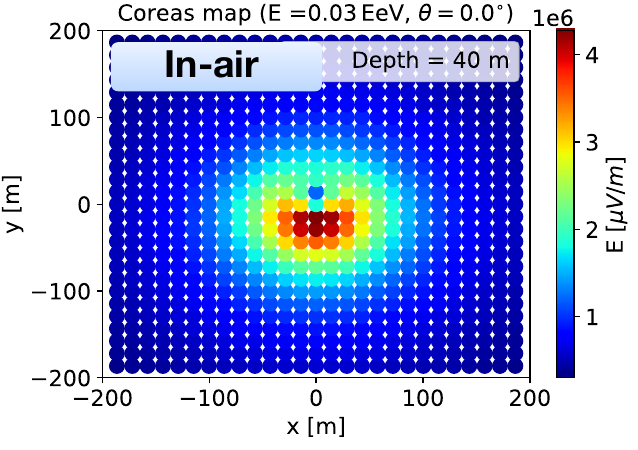}
    \end{minipage}
    \hfill
    \begin{minipage}[b]{0.49\textwidth}
        \centering
        \includegraphics[width=\textwidth]{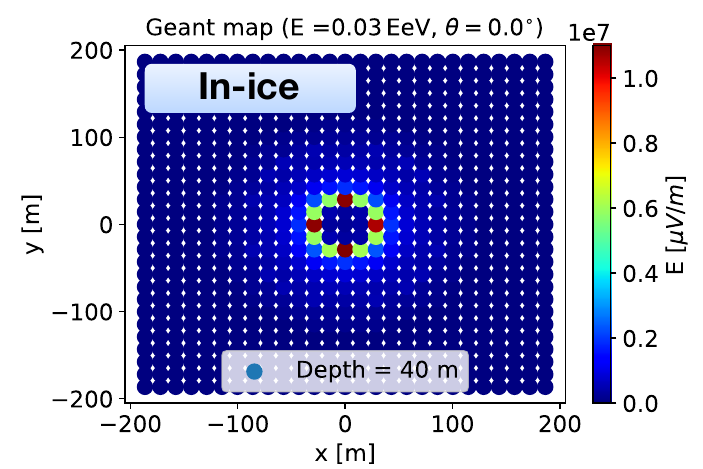}
    \end{minipage}

    \vspace{0.5cm}

    \begin{minipage}[b]{0.49\textwidth}
        \centering
        \includegraphics[width=\textwidth]{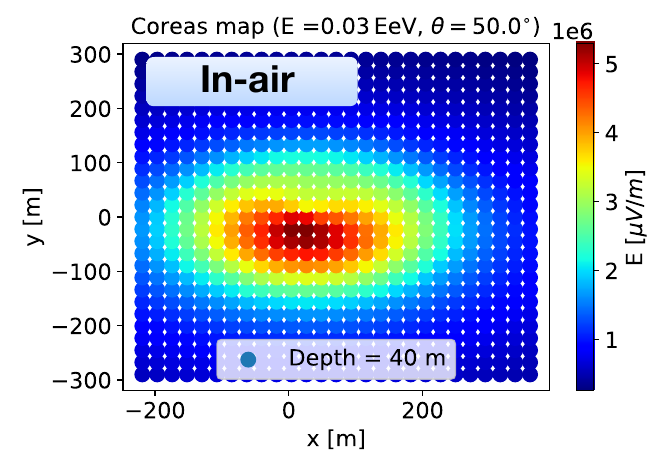}
    \end{minipage}
    \hfill
    \begin{minipage}[b]{0.49\textwidth}
        \centering
        \includegraphics[width=\textwidth]{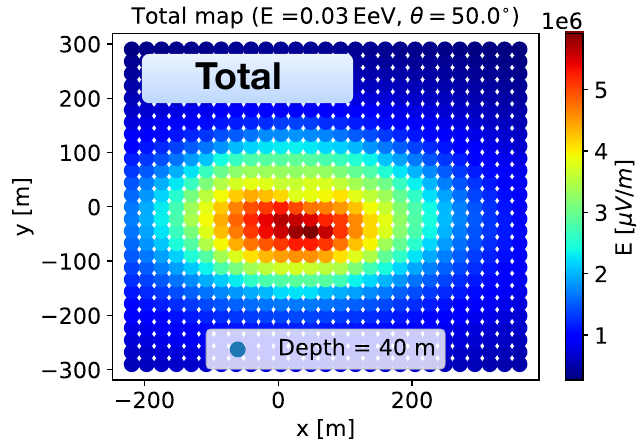}
    \end{minipage}
    
    \caption{{\it Top:} In-air ({\it left}) and in-ice ({\it right}) electric field maps of a vertical shower ($\theta  = 0^{\circ}$) at a depth $|z| = 40\, \rm m$. {\it Bottom:} In-air ({\it left}) and total ({\it right}) electric field maps of an inclined shower with $\theta  = 50^{\circ}$, at a depth $|z| = 40\, \rm m$. All the maps were simulated with the Greenland ice-model. }\label{fig:FluenceMaps}
\end{figure}

In Fig.~\ref{fig:FluenceMaps}, we show the electric field maps simulated with FAERIE for an antenna layer at a depth $|z| = 40{\, \rm m}$, corresponding to a shower with zenith angle $\theta = 0^{\circ}$ (top plots) and a shower with zenith angle $\theta = 50^{\circ}$ at the same depth (bottom plots).  For the top plots we separate between the emission from the in-air cascade (left-hand panel) and the one from the in-ice cascade (right-hand panel). For the in-air emission, we retrieve  a bean-shaped pattern as expected from the interference between the charge-excess and the geomagnetic emission. On the other hand, for the in-ice emission the radio signal is focused along an annulus-shaped region corresponding to the Cerenkov angle of the in-ice cascade. It should be noted that though the Cerenkov angle of the in-ice emission is larger than the one from the in-air emission, at a depth of $|z|=40\, {\rm m}$ both in-air and in-ice emissions are spread over a similar area, as the in-air emission point is located  much further away from the antenna grid. 

For the bottom plots (zenith angle $\theta = 50^{\circ}$),the left-hand plot shows a bean-shaped pattern for the in-air emission, similar to the previous case. However, this emission is spread over a larger area compared to the vertical shower, as inclined showers propagate through the atmosphere over longer distances than vertical ones and have an ellipsoidal projection on the ground. Finally, the bottom right-hand plot displays the coherent sum of the in-air and in-ice emissions at a zenith angle of $\theta = 50 ^{\circ}$. The resulting electric field map is very similar to that of the in-air emission alone. As mentioned in Section~\ref{sec:CRemission}, for inclined showers, only a few energetic particles reach the ground, so that the expected contribution to the total radio emission from the in-ice cascade is negligible.

\subsection{Cosmic-ray signatures}

 \begin{figure*}[tb]
\centering 
\includegraphics[width=0.49\columnwidth]{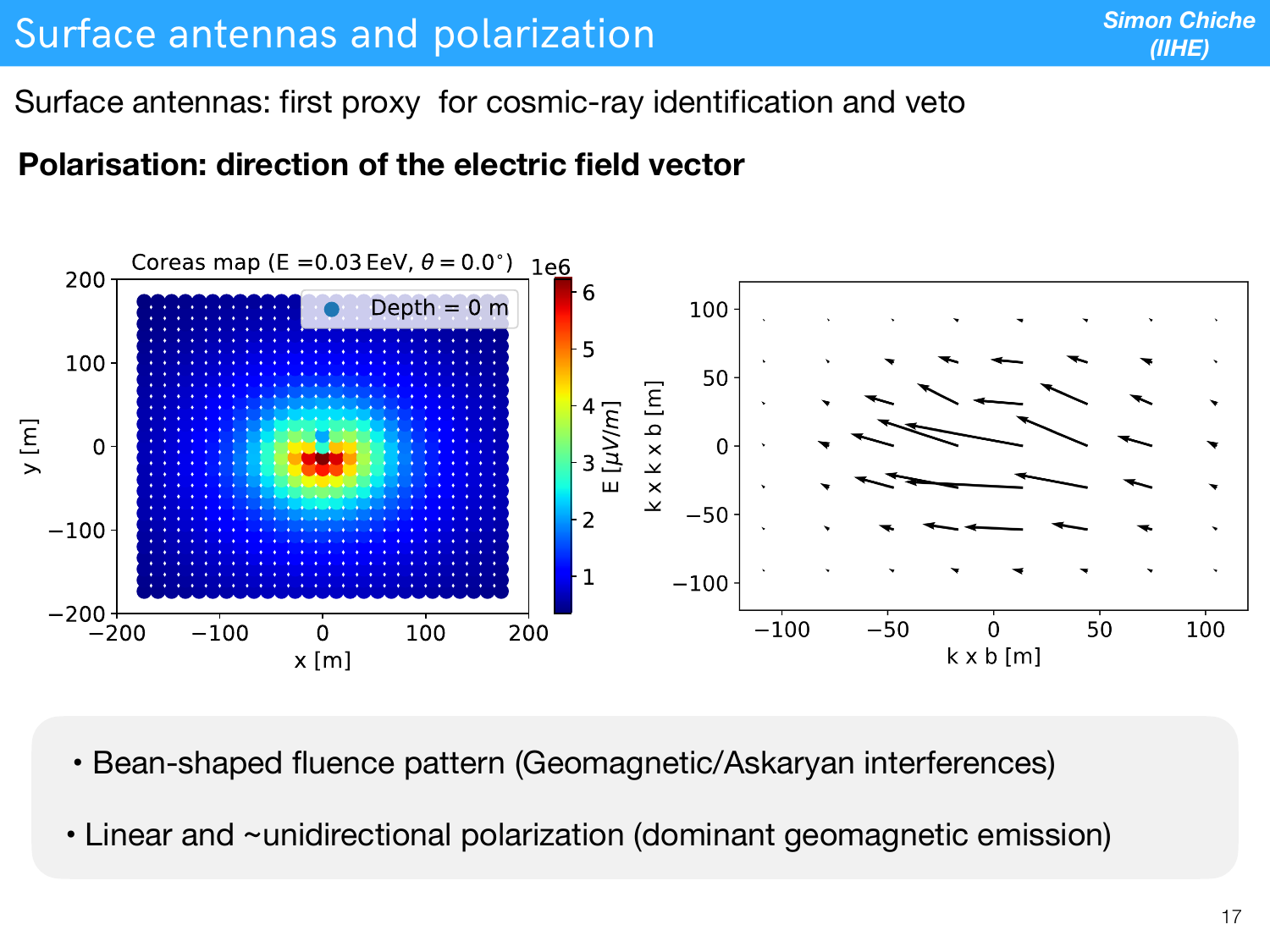}
\includegraphics[width=0.49\columnwidth]{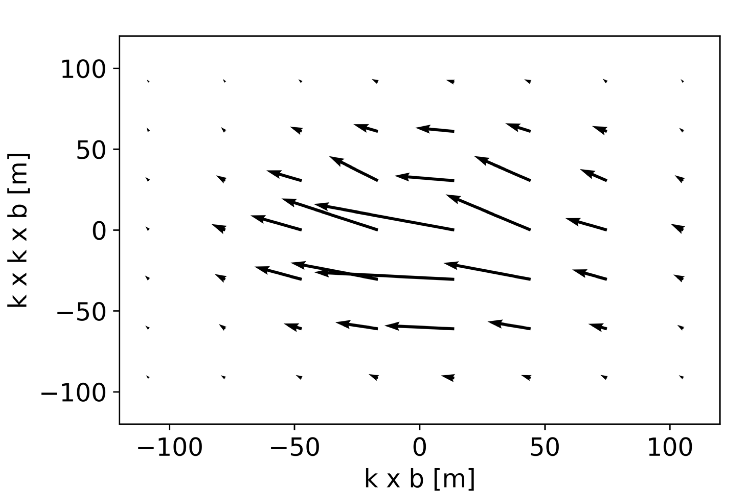}
\caption{{\it Left:} Simulated electric field map of a vertical shower at surface antennas (depth $|z| = 0\, \rm m$). {\it Right:}  Polarization vector of the electric field at the surface antennas. Each dot corresponds to an antenna and the length of the arrows is proportional to the amplitude of the electric field.
}\label{fig:Surface}
\end{figure*}

Using FAERIE simulations, we can then identify cosmic-ray signatures that would help us to design a discriminant between cosmic-rays and neutrinos. In Fig.~\ref{fig:Surface} we show the cosmic-ray induced emission measured by surface antennas, i.e., antennas located at the top of the ice sheet, as used in RNO-G stations. Since neutrino showers usually develop deep in the ice, no neutrino signal is expected at the surface antennas which should therefore act as a veto for cosmic-rays. This means that, the cosmic-ray induced emission detected by the surface antennas should come solely from the in-air cascade. In the left-hand panel of Fig.~\ref{fig:Surface}, we show that the electric field map of the total radio emission displays a bean-shaped pattern, as expected for an in-air emission.  In the right-hand panel of Fig.~\ref{fig:Surface}, we also show that the polarisation vector at the antenna level points dominantly towards the $\mathbf{-v \times B}$ direction, as expected from a dominant geomagnetic emission. 

\begin{figure*}[tb]
\centering 
\includegraphics[width=0.49\columnwidth]{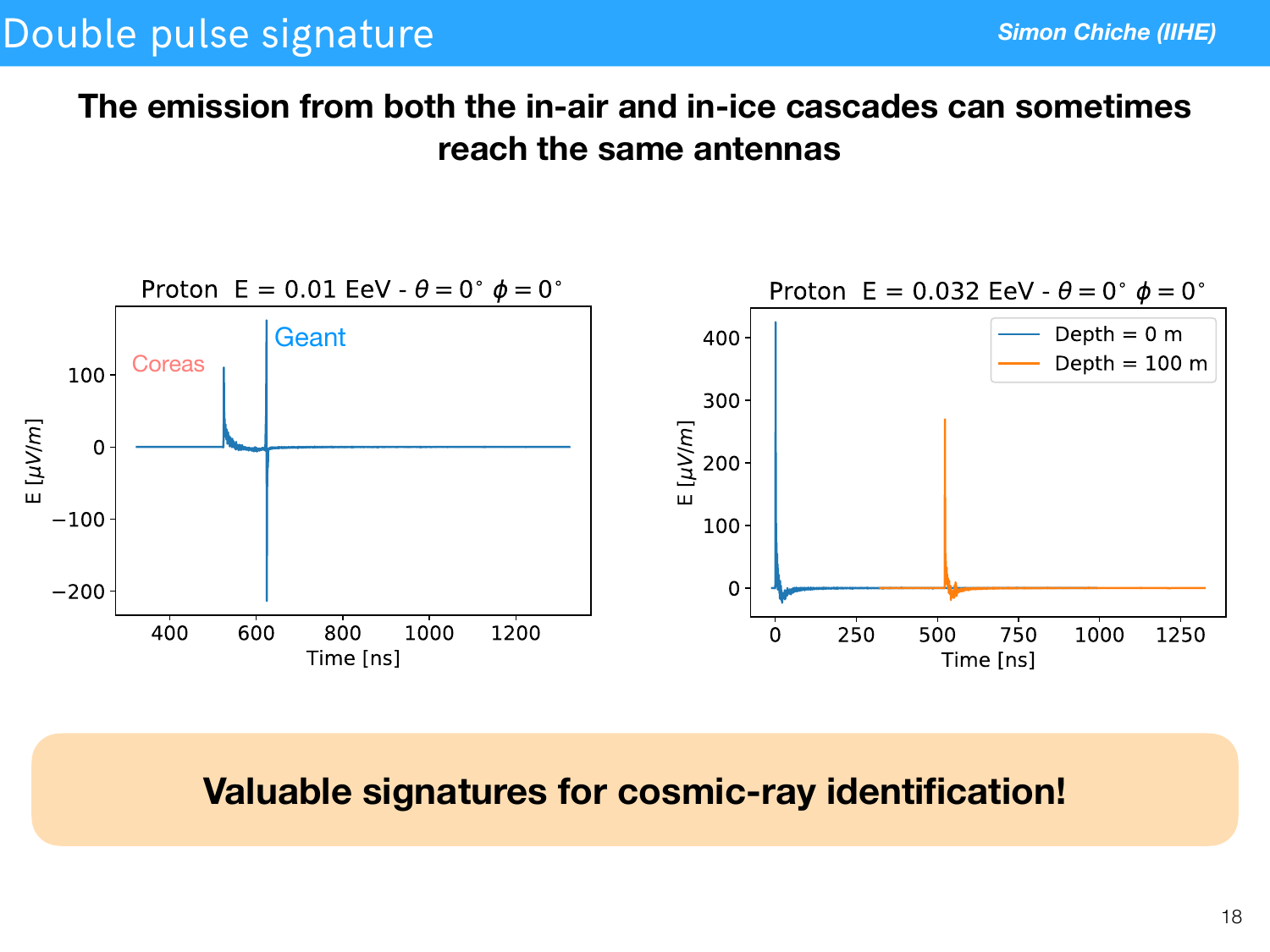}
\includegraphics[width=0.49\columnwidth]{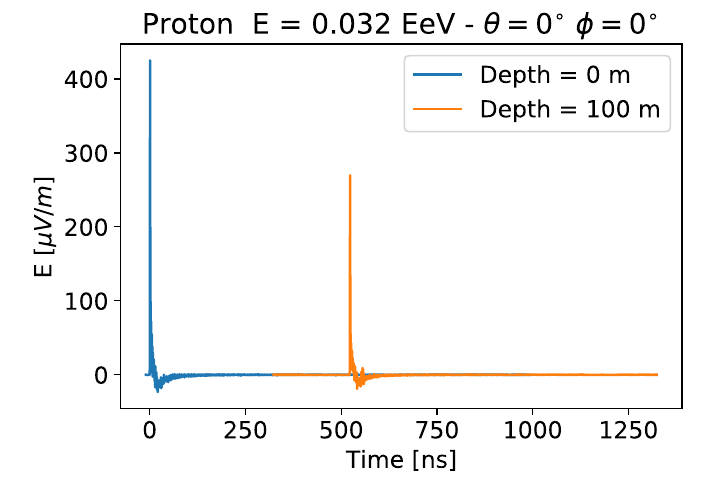}
\caption{ {\it Left:} Electric field time trace at a single antenna level with a double pulse signature. The first pulse is related to the emission from the in-air cascade (simulated with CoREAS), while the second pulse comes from the in-ice cascade (simulated with GEANT) and arrive with some delay.  {\it Right:}  Electric field time traces of an antenna at the ice's surface (blue pulse) and an antenna at a depth $|z| = 100\, \rm m$ located along the same line of sight, assuming $X_{\rm max}$ as emission point.
}\label{fig:Traces}
\end{figure*}

We can also use the electric field time traces to search  for cosmic-ray signatures. In the left-hand panel of Fig.~\ref{fig:Traces} we show a time trace at a given antenna with a typical double pulse signature. This feature is expected if both the in-air and the in-ice emission reach the same antenna and thus could be used to identify cosmic-ray induced showers. In the right-hand panel of Fig.~\ref{fig:Traces}, we also show the time trace of two antennas along the same line of sight, one antenna at a depth of $0\, \rm m$ (blue pulse) and the other at a depth of $100\, \rm m$ (orange pulse), considering $X_{\rm max}$ as emission point. We can see that the radio emission arrives earlier at the surface antenna and with a higher amplitude since this antenna is closer to $X_{\rm max}$. Additionally, we could link the delay between the two pulses to the path length of the radio emission between the antennas. Approximating this path length by the depth difference between the antennas, we get $\Delta t \sim 100\, \rm m \times n_{\rm ice} /c \sim 470 \, \rm ns$ if we assume $n_{\rm ice} =1.4$. These signatures should further help us to identify the cosmic-ray induced emission seen by in-ice observers.

\section{Conclusion}

Using the FAERIE Monte-Carlo code we simulated the radio emission induced by cosmic-ray showers deep in the ice, as targeted by experiments such as ARA and RNO-G. We studied the dependency of this emission with the shower geometry and, using raw traces, we identified preliminary cosmic-ray signatures based on the signal spatial distribution, polarization and timing. The successful detection of the first cosmic-ray event with deep antennas would be a breakthrough, as it would provide the first proof of concept for detecting particle cascades in nature using the in-ice radio technique. Such a detection would also allow in-ice experiments to calibrate their detectors and the identification of cosmic-ray signatures  will be crucial to ensure the design of a cosmic-ray/neutrino discriminant.


\newpage

\begin{thebibliography}{99}
\bibitem{Kockere_2024} S.D. Kockere {\it et al}, \Journal{Physical Review D}{110}{023010}{2024}.

\bibitem{Kravchenko_2003}I. Kravchenko {\it et al}, \Journal{Astroparticle Physics}{19}{15--36}{2003}.
\bibitem{Gorham_2009}P.W. Gorham {\it et al}, \Journal{Astroparticle Physics}{32}{10--41}{2009}.
\bibitem{Gorham_2019}P.W. Gorham {\it et al}, \Journal{Physical Review D}{99}{}{2019}.
\bibitem{Gorham_2021}P.W. Gorham {\it et al}, \Journal{Physical Review Letters}{126}{}{2021}.
\bibitem{Arianna}S. W. Barwick {\it et al}, \Journal{arxiv}{https://arxiv.org/abs/1410.7369}{}{2014}.
\bibitem{Anker_2019}A. Anker {\it et al}, \Journal{Advances in Space Research}{64}{2595--2609}{2019}.
\bibitem{Miller_2012}T. Miller {\it et al}, \Journal{Icarus}{220}{877--888}{2012}.
\bibitem{Allison_2016}P. Allison {\it et al}, \Journal{Physical Review D}{93}{}{2016}.
\bibitem{Aguilar_2021}J.A. Aguilar {\it et al}, \Journal{Journal of Instrumentation}{16}{P03025}{2021}.
\bibitem{Schr_der_2017}F. Schroeder, \Journal{Progress in Particle and Nuclear Physics}{93}{1--68}{2017}.
\bibitem{Heck:1998vt}D. Heck {\it et al}, \Journal{CORSIKA: a Monte Carlo code to simulate extensive air showers}{}{}{1998}.
\bibitem{ALLISON2016186}J. Allison {\it et al}, \Journal{Nuclear Instruments and Methods in Physics Research}{835}{186-225}{2016}.
\bibitem{HuegeCOREAS}T. Huege {\it et al}, \Journal{AIP Conference Proceedings}{1535}{128-132}{2013}.
\bibitem{T510}K. Bechtol {\it et al}, \Journal{Phys. Rev. D}{105}{063025}{2022}.
\bibitem{PhysRevE.84.056602}C.W. James {\it et al}, \Journal{Phys. Rev. E}{84}{056602}{2011}.
\bibitem{Kelley:2017ws}J. Kelley {\it et al}, \Journal{PoS}{ICRC2017}{1030}{2017}.
\bibitem{RNOice}C. Deaconu {\it et al}, \Journal{Physical Review D}{98}{043010}{2018}.
\end{thebibliography}
\end{document}